\documentclass[sigconf]{acmart}
\AtBeginDocument{%
  \providecommand\BibTeX{{%
    \normalfont B\kern-0.5em{\scshape i\kern-0.25em b}\kern-0.8em\TeX}}}

\copyrightyear{2021}
\acmYear{2021}

\acmConference[ORSUM@ACM RecSys 2021]{ORSUM@ACM RecSys 2021}{October 2nd, 2021}{Amsterdam, the Netherlands}
\acmBooktitle{ORSUM@ACM RecSys 2021, Amsterdam, the Netherlands}

\theoremstyle{definition}

\usepackage{listings}

\usepackage{amsmath}
\usepackage{mathtools}
\DeclarePairedDelimiter{\ceil}{\lceil}{\rceil}
\usepackage{multirow}

\usepackage{subcaption}
\usepackage{float}

\usepackage{enumitem}

\newcommand{\header}[1]{\vspace{1mm}\noindent\textbf{#1}.}
\newcommand{\headerl}[1]{\vspace{1mm}\noindent\textit{#1}.}

\usepackage{algorithm}
\usepackage[noend]{algpseudocode}
\usepackage{listings}
\usepackage{xcolor}

\lstdefinelanguage{scala}{
  morekeywords={abstract,case,catch,class,def,%
    do,else,extends,false,final,finally,%
    for,if,implicit,import,match,mixin,%
    new,null,object,override,package,%
    private,protected,requires,return,sealed,%
    super,this,throw,trait,true,try,%
    type,val,var,while,with,yield},
  otherkeywords={=>,<-,<\%,<:,>:,\#,@},
  sensitive=true,
  morecomment=[l]{//},
  morecomment=[n]{/*}{*/},
  morestring=[b]",
  morestring=[b]',
  morestring=[b]"""
}

\begin{document}

\title[Efficiently Maintaining Next Basket Recommendations]{Efficiently Maintaining Next Basket Recommendations under~Additions and Deletions of Baskets and Items}

\author{Benjamin Longxiang Wang}
\affiliation{%
  \institution{University of Amsterdam \& ABN AMRO Bank}
  \country{}
  }
\email{wlongxiang1119@gmail.com}

\author{Sebastian Schelter}
\affiliation{%
  \institution{AIRLab, University of Amsterdam}
  \country{}
  }
\email{s.schelter@uva.nl}

\renewcommand{\shortauthors}{}


\begin{abstract}

Recommender systems play an important role in helping people find information and make decisions in today's increasingly digitalized societies. However, the wide adoption of such machine learning applications also causes concerns in terms of data privacy. These concerns are addressed by the recent ``General Data Protection Regulation'' (GDPR) in Europe, which requires companies to delete personal user data upon request, when users enforce their ``right to be forgotten''. Many researchers argue that this deletion obligation does not only apply to the data stored in primary data stores such as relational databases, but also requires an update of machine learning models whose trainingset included the personal data to delete. As a consequence, the academic community started to investigate how to unlearn user data from trained machine learning models in an efficient and timely manner. 

We explore this direction in the context of a sequential recommendation task called Next Basket Recommendation (NBR), where the goal is to recommend a set of items based on a user's purchase history. We design efficient algorithms for incrementally and decrementally updating a state-of-the-art next basket recommendation model in response to additions and deletions of user baskets and items. Furthermore, we discuss an efficient, data-parallel implementation of our method in the Spark Structured Streaming system. 

We evaluate our implementation on a variety of real-world datasets, where we investigate the impact of our update techniques on several ranking metrics and measure the time to perform model updates. Our results show that our method provides constant update time efficiency with respect to an additional  user basket in the incremental case, and linear efficiency in the decremental case where we delete existing baskets. With modest computational resources, we are able to update models with a latency of around 0.2~milliseconds regardless of the history size in the incremental case, and less than one millisecond in the decremental case.
\end{abstract}

\maketitle

\section{Introduction}

Recommender systems (RS) are ubiquitous in our digitalized societies with applications in e-commerce, social networks and streaming services~\cite{sun2019research}.  However, the wide adoption of such machine learning applications has been causing concerns in terms of fairness, accountability, privacy, and transparency in the general public and research community alike~\cite{Sinha2002transparency, zhang2014privacy, burke2018fairness}. It is widely recognised that we need to tackle such responsibility issues at the intersection of machine learning and data management~\cite{stoyanovich2020responsible,schelter2018challenges,schelter2015efficient,kunft2017blockjoin}.

\header{The right to be forgotten} In this work, we focus on one specific aspect of privacy stipulated by the \textit{General Data Protection Regulation} (GDPR), namely, ``the right to be forgotten''~\cite{gdpr-forgotten-right-article-17}, which states that the \textit{``data subject shall have the right to [...] the erasure of personal data concerning him or her without undue delay [...] where the data subject withdraws consent''}. When users can enforce their right to be forgotten, companies must  delete their personal user data. Many researchers argue that this deletion obligation does not only apply to the data stored in primary data stores such as relational databases, but also requires an update of machine learning models whose trainingset included the personal data to delete. Legal scholars even point out the continual use of AI systems trained on deleted data could be considered illegal under certain interpretations~\cite{legalgdprhumanforgetsairemembers}. In some cases, the retained lossy representations in ML models can even be restored by adversarial parties \cite{Schelter2020AmnesiaM}.

\header{Towards timely and efficient unlearning} A naive way to realize data deletion is to retrain the model from scratch on the remaining dataset, leaving out the user data to delete. However, this approach is typically expensive in terms of time and computational resources. Furthermore, the GDPR mentions companies should erase data \textit{``without undue delay''} \cite{gdpr-forgotten-right-article-17} but does not directly specify when deletions should be fully conducted. In practice, data deletion is still a lengthy process. For example, Google could take two to six months to fully delete someone's data \footnote{\url{https://policies.google.com/technologies/retention}}; for Facebook, this period is 90 days \footnote{\url{https://www.facebook.com/help/250563911970368}}. We argue that it is an open and important academic question to determine how fast data can be erased from recommendation models, and that \textit{we should design algorithms and systems to empower users to exercise their right to be forgotten as timely as possible}, instead of making them wait for several months.

\header{Efficiently maintaining a TIFU-kNN model for next basket recommendation} In this paper, we explore the direction of efficiently updating a recommendation model in the context of a sequential recommendation task called ``Next Basket Recommendation''~(NBR, \autoref{sec:background-nbr}), where the goal is to recommend a set of items based on a user's purchase history. We focus our efforts on \textit{TIFU-kNN} (Temporal Item Frequency-based User-kNN)~\cite{tifuknn}, a nearest neighbor-based  state-of-the-art next basket recommendation model~(\autoref{sec:tifuknn}), which outperforms neural networks in the NBR task, despite its simplicity. We design efficient algorithms for incrementally and decrementally updating a state-of-the-art next basket recommendation model in response to additions and deletions of user baskets and items.\\

We formalize the problem of maintaining a TIFU-kNN model under  additions and deletions of baskets and items in \autoref{sec:problem_statement}. Next, we design efficient online algorithms that extend TIFU-kNN to learn incrementally as new data arrives (incremental updates \autoref{sec:incrementalapproach}) and to forget efficiently as deletion requests come in (decremental updates \autoref{sec:decrementalapproach}). We implement our approach in a data-parallel manner in the Apache Spark Structured Streaming system~\cite{zaharia2016apachespark} (\autoref{sec:system_overview}). Furthermore, we conduct extensive experiments in \autoref{section:evaluation} where we analyze the recommendation performance and update efficiency of our approach, and point out its limitations for an extreme number of deletions in \autoref{section:limitations}. In summary, this paper provides the following contributions.

\begin{itemize}[leftmargin=*]
    \item We introduce the problem of incremental/decremental learning for next basket recommendation~(\autoref{sec:problem_statement}).
    \item We design efficient online algorithms for incremental and decremental updates of the state-of-the-art NBR model TIFU-kNN~(\autoref{sec:model_overview}).
    \item We discuss a data-parallel implementation of our approach in the Spark Structured Streaming system~(\autoref{sec:system_overview}).
    \item We conduct experiments on three real-world datasets, where we evaluate our approach in terms of predictive performance and update time efficiency. We find that we are able to update the model with constant time efficiency in the incremental case and linear time efficiency in the decremental case with millisecond latency (\autoref{section:evaluation}).   
\end{itemize}
\section{Background}
\label{sec:background}

\subsection{Next Basket Recommendation}
\label{sec:background-nbr}

We focus on a specific recommendation task called Next Basket Recommendation (NBR), where we aim to recommend a set of items to users based on their historical baskets~\cite{rendle2010FPMCforNBR}. NBR is of great interest to the e-commerce and retail industry, where we want to recommend a set of items to fill a user's shopping basket. 
A basket $\boldsymbol{b}$ is a set of items, i.e., $\boldsymbol{b} = \{i_1, i_2,\cdots, i_j,\cdots, i_{|\boldsymbol{b}|}\}$, where $i_j \in \mathcal{I}$, and where $\mathcal{I}$ denotes the universe of all items. For a given user, we have access to a sequence of $n$ historical baskets (in increasing chronological order, such that more recent items are at the tail) denoted as $\mathcal{H} = [\boldsymbol{b_1},\boldsymbol{b_2},\cdots,\boldsymbol{b_i},\cdots,\boldsymbol{b_n}]$, where $\boldsymbol{b_i} \subset \mathcal{I}$. The goal of NBR is then to a find a model which takes the historical baskets $\mathcal{H}$ as input and predicts the next basket $\boldsymbol{b_{n+1}}$ as recommendation.

\subsection{TIFU-kNN}
\label{sec:tifuknn}
Our work centers around adapting the state-of-the-art next basket recommendation model TIFU-kNN~\cite{tifuknn}. This method builds on the user-based collaborative filtering approach \cite{collaborativefiltering}. A target user is first modeled by an embedding vector, and then recommendations are made based on a weighted combination of the aggregated representations of similar users and the target user's own vector embedding.  TIFU-kNN calculates user vector representations during the training phase leveraging repeated purchase patterns \cite{repeatdynamics, repeatdynamics2} commonly observed in e-commerce, under the assumption that recent purchases have more predictive power than older ones.

\header{User vector representation} A user vector is calculated to represent a user's past purchase patterns, with a time-decayed window to assign a higher weight to recent items. However, a single time decayed weight is not flexible enough to model another property of temporal dynamics: consecutive steps have small changes while steps far from each other have large changes. Therefore, TIFU-kNN applies a hierarchical structure where user baskets are first divided into groups, baskets within a group are aggregated with time decayed weights to obtain a group vector, and afterwards, these group vectors are aggregated again with time decayed weights to obtain a final user vector. In detail, the algorithm applies the following steps:

\headerl{Step 1: Multi-hot encoding} Without loss of generality, we denote the historical baskets of our target user as $\mathcal{H} = [\boldsymbol{b_1},\cdots,\boldsymbol{b_i},\cdots,\boldsymbol{b_n}]$ where $\boldsymbol{b_i} \subset \mathcal{I}$ and $|\mathcal{H}| = n$. Each basket $\boldsymbol{b_i}$ is converted into a multi-hot encoded vector whose dimension is equal to $|\mathcal{I}|$ with ones in dimensions for items present in the basket, and zeros in the remaining positions. For example, suppose we have a corpus containing four items, i.e., $\mathcal{I}= \{1, 2,3,4\}$, and a user's history with two baskets $\mathcal{H} = [\{1,4\}, \{1, 2,3\}]$. The corresponding multi-hot vector would then look as follows.
\begin{align*}
    [\{1,4\}, \{1, 2,3\}] \xRightarrow[ \{1, 2,3,4\} ] {\text{multi-hot encoding}} [ [1, 0, 0, 1], [1, 1, 1, 0]]
\end{align*}

\headerl{Step 2: Group vector generation} To model the user's behavior shift over time, TIFU-kNN utilizes a hierarchical grouping mechanism to capture the dynamics. Given a fixed group size $m$ as a hyper-parameter, the number of groups is denoted as $ k = \ceil*{\frac{n}{m}}$ where $\ceil*{.}$ is the ceiling function. The group size dictates the number of baskets in a group. Given a temporal decay rate $0<r_b \le 1 $ and baskets partitioned into groups of equal length (except for the last group which contains the remaining baskets in case $n$ is not fully divisible by $m$), a group division example looks as follows.
\begin{align*}
\mathcal{H} &= [\overbrace{\boldsymbol{b}_1,\cdots, \boldsymbol{b}_m}^{\boldsymbol{v_{g1}}},\overbrace{\boldsymbol{b}_{m+1},\cdots, \boldsymbol{b}_{2m}}^{\boldsymbol{v_{g2}}},...,\overbrace{\cdots,\boldsymbol{b}_{n-1}, \boldsymbol{b}_n}^{\boldsymbol{v_{gk}}} ]
\end{align*}

Each group is represented by a group vector $\boldsymbol{v_{g}}$ which is calculated by a time-decay weighted average, where the \textit{i-th} vector within a group is multiplied by a time-decayed weight $r_b^{m-i}$, where $\boldsymbol{v_{bi}}$ is a multi-hot encoded vector. Note that the indicator for the group index is left out for brevity.
\begin{align}
\boldsymbol{v_{g}} =\frac{r_b^{m-1}\boldsymbol{v_{b1}} + r_b^{m-2}\boldsymbol{v_{b2}} + \cdots+ \boldsymbol{v_{bm}} }{m} = \frac{\sum_{i=1}^m r_b^{m-i}\boldsymbol{v_{bi}} }{m}
\label{eq: groupvector}
\end{align}

\headerl{Step 3: User vector generation}
. Similar to how a group vector is calculated from basket vectors, TIFU-kNN calculates the final target user vector $\boldsymbol{v_{u}}$representation based on a time-decay weighted average of the group vectors $\mathcal{H} = [\boldsymbol{v_{g1}}, \boldsymbol{v_{g2}, \cdots,\boldsymbol{v_{gk}}}]$. Given the group decay rate $0<r_g \le 1 $, we have
\begin{align} 
    \boldsymbol{v_{u}} = \frac{ r_g^{k-1}\boldsymbol{v_{g1}} + r_g^{k-2}\boldsymbol{v_{g2}} + \cdots +  \boldsymbol{v_{gk}}}{k} =  \frac{\sum_{i=1}^k r_g^{k-i}\boldsymbol{v_{gi}} }{k}  \label{eq: uservector}
\end{align}

\header{Personalized Collaborative Filtering Recommendation} After obtaining the vector representations of all users, TIFU-kNN generates recommendations inspired by collaborative filtering~(CF)~\cite{collaborativefiltering}. Given a predefined distance measure, it finds the top-$k$ most similar users for each target user. The final prediction combines two parts: $(i)$~the personalized component, e.g., the user vector representation denoted by $\boldsymbol{u_{\text{target}}}$, corresponding to the user's temporal repeated purchase pattern; $(ii)$~the collaborative filtering component, which is calculated as the mean of the nearest neighbors' vector representations, denoted by $\boldsymbol{u_{\text{neighbors}}}$. 

The final prediction vector is calculated via a linear combination $\boldsymbol{p} = \alpha \boldsymbol{u_{\text{target}}} + (1- \alpha) \boldsymbol{u_{\text{neighbors}}}$ of both components, where $\alpha$ is a hyper-parameter used to adjust the weight of each component. Note that the numeric values in the prediction vector $\boldsymbol{p}$ indicate the strength of preference of the user towards a certain item $\boldsymbol{p}$, and the highest scored items are typically leveraged for recommendation.
\section{Problem Statement}
\label{sec:problem_statement}

As outlined in the introduction, we focus on two aspects of the maintenance of TIFU-kNN models: $(i)$~how to incrementally learn user representations in response to additional baskets~(\textit{incremental updates}); $(ii)$~how to efficiently delete user data from the learned user representations~(\textit{decremental updates}). 

Let $\mathcal{D} = \{\mathcal{H}_1,\mathcal{H}_2,\cdots,\mathcal{H}_i, \cdots,\mathcal{H}_M \}$ denote a dataset containing historical baskets for $M$ users, and the corresponding user vectors are $\mathcal{U} = \{\boldsymbol{v_{u1}},\boldsymbol{v_{u2}}, \cdots,\boldsymbol{v_{ui}}, \cdots, \boldsymbol{v_{uM}} \}$ where $\boldsymbol{v_{ui}}$ is a real-valued vector of size $|\mathcal{I}|$. Our goal is to maintain the user vector of a TIFU-kNN model. As each user vector is modeled independently (e.g., only calculated from the baskets of the particular user), we consider only a single user in the following.

\header{Constant time incremental updates} Given the historical baskets $\mathcal{H} = [\boldsymbol{b_1},\boldsymbol{b_2},\cdots,\boldsymbol{b_i},\cdots,\boldsymbol{b_n}]$ and a newly arriving basket $\boldsymbol{b_{n+1}}$ for user $\boldsymbol{u}$, we can always recalculate the corresponding user vector from scratch from the complete history. For incremental updates, we want to improve the asymptotic time complexity from $\mathcal{O}(|\mathcal{H}|)$ for retraining from scratch to a constant effort $\mathcal{O}(1)$ per additional basket. Given the history $\mathcal{H}$ and the current user vector $\boldsymbol{v_u}$, we are interested in an efficient approach $f_{\text{incr}} (\boldsymbol{v_u},\boldsymbol{b_{n+1}}) \rightarrow \boldsymbol{v_u^\prime}$ to update the current user vector $\boldsymbol{v_u}$ to $\boldsymbol{v_u^\prime}$ by incorporating $\boldsymbol{b_{n+1}}$.

\header{Efficient decremental updates} In the decremental case, given the current user vector $\boldsymbol{v_u}$ and a deletion request $r$ (identifying a basket or item to be deleted), we aim to efficiently delete $r$ from the corresponding user vector $\boldsymbol{v_u}$. We want to improve the asymptotic complexity from $\mathcal{O}(|\mathcal{H}|)$ for retraining from scratch to $\mathcal{O}(|\mathcal{H}| - p)$ where $p$ is the position of the basket where a deletion is applied. Formally, we look for a decremental update function $f_{\text{decr}}(\mathcal{H}, \boldsymbol{v_u}, r)$ which takes three inputs: the historical baskets $\mathcal{H}$, the existing user vector $\boldsymbol{v_u}$ and the deletion request $r$, and outputs an updated user vector $\boldsymbol{v_u^\prime}$. We want $f_{\text{decr}}$ to satisfy $\forall \boldsymbol{u}, \forall r: f_{\text{decr}}(\mathcal{H}, \boldsymbol{v_u}, r) = \boldsymbol{v_u^\prime} \label{eq:exact}$, e.g., we want the result of a decremental update to be the same as computing the user vector after deletion $\boldsymbol{v_u^\prime}$ by running TIFU-kNN from scratch on $\mathcal{H} \setminus r$.
\section{Approach}
\label{sec:model_overview}

As mentioned in \autoref{sec:problem_statement}, we compare our approach to the baseline approach of retraining the whole model from scratch whenever there is an incremental or decremental update. This baseline has an asymptotic complexity of $\mathcal{O}(|\mathcal{H}|)$ per update, be it incremental or decremental. This baseline has several disadvantages: $(i)$~the (re)training phase is expensive in terms of time and resources, especially if it has to be executed repeatedly for a small number of data deletion requests $(ii)$ the user has to potentially wait long for their data to be actually deleted from the model if the retraining is only executed periodically. 

To overcome these challenges, we design online algorithms which update the model with low latency in response to additional data or deletion requests concerning baskets or items. We first discuss general rules for maintaining the decaying average of a series in \autoref{sec:das} and then apply them to derive our approach to incremental updates in \autoref{sec:incrementalapproach} and decremental updates in \autoref{sec:decrementalapproach} under different scenarios.

\subsection{Maintaining the Decaying Average of a Series}
\label{sec:das}
Calculating decaying averages of vector series is at the heart of TIFU-kNN (\autoref{sec:tifuknn}) as it is the core computation for both the group vector (\autoref{eq: groupvector}) and user vector (\autoref{eq: uservector}). Without loss of generality, we discuss this problem with a series of real numbers. Let $\mathcal{S} = [x_1,\cdots, x_n] \label{eq: streamdef}$ denote a series of real numbers, where the next item is $x_{n+1}$. Let $\Bar{x}_{n}$ denote the decaying average of the first $n$ numbers. Given  $0<r \le 1$, the decaying average of the series is $\Bar{x}_{n}=\frac{1}{n} \sum_{i=1}^{n} r^{n-i} x_i$.
We examine how the decaying average changes in response to incremental and decremental updates. Note that it is straightforward to extend our approach to a series of vectors of real numbers by replacing the enclosed elements by vectors of the same length, which essentially leads to the group vector (\autoref{eq: groupvector}) and user vector (\autoref{eq: uservector}) definitions of TIFU-kNN. We now look at three scenarios: incremental updates, decremental updates, and in-place updates.

\header{Incrementally adding a number} With a newly arriving item $x_{n+1}$, we update the corresponding decaying average as follows.
\begin{align} 
    \Bar{x}_{n+1}  &= \frac{1}{n+1} \sum_{i=1}^{n} r^{n+1-i} x_i \nonumber \\
    &=  \frac{r^{n} x_1 + r^{n-1} x_2 + ...+ rx_{n} + x_{n+1}}{n+1} \nonumber \\
    &=  \frac{rn\Bar{x}_{n} + x_{n+1}}{n+1} \label{eq: inc_dsa}
\end{align}
With this update rule, we can efficiently update the decaying average in constant time using only the current decaying average $\Bar{x}_{n}$, the count of items $n$, the additional item $x_{n+1}$ and the decay rate $r$.

\header{Decrementally deleting a number} If an existing number $x_i$ needs to be deleted from the series, we would like to find an efficient way to update the decaying average without resorting to a full recalculation. Before deletion, the decaying average is:
\begin{align*}
    \Bar{x}_{n}=\frac{\overbrace{ r^{n-1}x_1 + \cdots + r^{n-(i-1)}x_{i-1}}^{preceding\ deletion} +  \overbrace{r^{n-i}x_i}^{deletion} + \overbrace{r^{n-(i+1)}x_{i+1} +\cdots + x_n}^{succeeding\ deletion}}{n}
\end{align*}

After deletion, the updated decaying average becomes:
\begin{align*}
    \Bar{x}_{n-1}^\prime
    &= \frac{\overbrace{ r^{n-2}x_1 + \cdots + r^{n-1-(i-1)}x_{i-1}}^{preceding\ deletion}  + \overbrace{r^{n-(i+1)}x_{i+1} +\cdots + x_n}^{succeeding\ deletion}}{n-1} 
\end{align*}

We rearrange the terms such that we express $\Bar{x}_{n-1}^\prime$ using $\Bar{x}_{n}$:
\begin{align} 
    \Bar{x}_{n-1}^\prime
    &= \frac{n\Bar{x}_{n} + [r^{n-i} (x_{i+1} - x_i) + \cdots +  r (x_{n} - x_{n-1}) - x_n]}{(n-1)r} \nonumber\\
    &= \frac{n\Bar{x}_{n} + \mathcal{D}([x_{i}, \cdots,x_n])^T   \mathcal{R}(r,n-i)}{(n-1)r}\label{eq: dec_dsa}
\end{align}
where $\mathcal{D}(.)$ denotes a vector of the first order difference of the slice of the series \textit{after} the deleted item:
$\mathcal{D}([x_{i}, \cdots,x_n]) = [x_{i+1} - x_i, \cdots, x_n - x_{n-1}, -x_n]^T$ and $\mathcal{R}(.)$ denotes a decaying vector: $\mathcal{R}(r,n-i) = [r^{n-i}, \cdots, r, 1]^T$. We discuss a few observations:
\begin{itemize}[leftmargin=*]
    \item The decremental update rule (\autoref{eq: dec_dsa}) is mathematically symmetrical to its incremental counterpart (\autoref{eq: inc_dsa}), as in both cases we update by incorporating the previous value and the change introduced by the additional or removed item. This view provides a unified perspective to deal with incremental and decremental updates.
    \item  To perform a deletion, it is sufficient to access a slice starting from the deleted item. In the worst case when the deleted item is the very first item $x_1$, the slice is equivalent to the full series; in the best case when the deleted item is the very last item, we only need to access a single item. Compared to a full recalculation where (re)access to the full history is needed, our update rule reduces the amortized number of items impacted by a deletion from $|\mathcal{S}|$ to $|\mathcal{S}|/2$, assuming a uniform distribution of deletions.
    \item Lastly the term $\mathcal{D}(.)^T \mathcal{R}(.)$ is a dot product and can be efficiently computed by vectorization, which further improves the efficiency of decremental updates.
\end{itemize}

\header{Updating an existing number} Suppose $x_i$ is updated to a new value $x_i^\prime$, then the new decaying average becomes
\begin{align} 
    \Bar{x}_{n}^\prime
    &= \frac{r^{n-1}x_1 + \cdots + r^{n-i}x_i^\prime +\cdots + x_n}{n} \nonumber\\
    &=\frac{r^{n-1}x_1 + \cdots + r^{n-i}x_i^\prime +\cdots + x_n +  r^{n-i}x_i -  r^{n-i}x_i}{n} \nonumber\\
    &= \frac{ n\Bar{x}_{n} +   r^{n-i}(x_i^\prime - x_i)}{n} \label{eq: inplace_dsa}
\end{align}
Note that for an in-place update, we only need to update the previous decaying average with a term dependent on the difference introduced by the updated item.

The approaches to update decaying averages for the incremental, decremental, and in-place scenarios serve as a basis for the online algorithms for TIFU-kNN developed in the following subsections.

\subsection{Incrementally Adding New Baskets}
\label{sec:incrementalapproach}
Given an existing user vector $\boldsymbol{v_u}$ and an additional basket $\boldsymbol{b_{n+1}}$, we discuss an algorithm to update the user vector efficiently based on the rules discussed in \autoref{sec:das}. Note that it is sufficient just to study the problem for a single user as all user vectors are computed independently.
Due to the grouping mechanism in TIFU-kNN~(\autoref{sec:background}), adding a basket to the history will either lead to 1) a new group with a single basket (if the last existing group is fully occupied) or 2) an update of the last existing group with one new basket appended. This is best illustrated with an example. Consider a user with four baskets, group size $m = 2$, hence number of groups $k=\ceil*{\frac{4}{2}} = 2$ which leads to two groups with each group containing two baskets. Now we add two new baskets one by one. The group composition evolves as follows.
\begin{align}
             &\mathcal{H} = [\overbrace{\boldsymbol{v_{b1}},\boldsymbol{v_{b2}}}^{\boldsymbol{v_{g1}}},\overbrace{\boldsymbol{v_{b3}},\boldsymbol{v_{b4}}}^{\boldsymbol{v_{g2}}}] \rightarrow \nonumber\\
             &\mathcal{H}^\prime = [\overbrace{\boldsymbol{v_{b1}},\boldsymbol{v_{b2}}}^{\boldsymbol{v_{g1}}},\overbrace{\boldsymbol{v_{b3}},\boldsymbol{v_{b4}}}^{\boldsymbol{v_{g2}}},\overbrace{\boldsymbol{v_{b5}}}^{\boldsymbol{v_{g3}}} ] \rightarrow \nonumber\\
             &\mathcal{H}^{\prime\prime} = [\overbrace{\boldsymbol{v_{b1}},\boldsymbol{v_{b2}}}^{\boldsymbol{v_{g1}}},\overbrace{\boldsymbol{v_{b3}},\boldsymbol{v_{b4}}}^{\boldsymbol{v_{g2}}},\overbrace{\boldsymbol{v_{b5}}, \boldsymbol{v_{b6}}}^{\boldsymbol{v_{g3}^\prime}} ]
             \label{eq:example}
\end{align}
A group with a single basket is created after adding a new basket to a fully occupied group, and afterwards, the group is updated with the second additional basket. Our update mechanisms differentiate these two scenarios:

\header{Scenario 1: Adding a group with a single basket}
When transitioning from $\mathcal{H}$ to $\mathcal{H}^\prime$ (\autoref{eq:example}), we add an additional  group vector (with a single basket) to update the user vector while the existing group vectors remain unchanged. Recall that we work with a group size $m$, a number of baskets $n$ and a number of groups $k = \ceil*{\frac{n}{m}}$. Since the new group only contains the newly added basket, the new group vector is equivalent to the new basket vector $\boldsymbol{v_{g(k+1)}} = \boldsymbol{v_{b(n+1)}}  \label{eq: inc_group_vector_rule_1}$. Given the additional group vector $\boldsymbol{v_{g(k+1)}}$, the existing user vector $\boldsymbol{v_{u}}$ as defined in \autoref{eq: uservector}, and the group decay rate $r_g$, we obtain the updated user vector $\boldsymbol{v_{u}^\prime}$ by applying the incremental rule derived in \autoref{eq: inc_dsa}.
\begin{align}
    \boldsymbol{v_{u}^\prime} &= \frac{r_g^{k}\boldsymbol{v_{g1}} + r_g^{k-1}\boldsymbol{v_{g2}} + \cdots + r_g\boldsymbol{v_{gk}} + \boldsymbol{v_{g(k+1)}}  }{k+1} \nonumber\\
    &= \frac{kr_g\boldsymbol{v_{u}}+ \boldsymbol{v_{b(n+1)}}}{k+1} \label{eq: inc_user_vector_rule_1}
\end{align}

Note that we can efficiently update the user vector in constant time regardless of the size of the user's history with access to the current user vector $\boldsymbol{v_{u}}$ and the number of groups $k$.

\header{Scenario 2: Updating an existing group}
When transitioning from $\mathcal{H}^\prime$ to $\mathcal{H}^{\prime\prime}$ (\autoref{eq:example}), we have $\tau < m$ where $\tau$ is the number of baskets in the last group and $m$ is the group size. In this scenario, the last group vector $\boldsymbol{v_{gk}}$ needs to be updated to take the added basket into account. Given the number of existing baskets in the last group $\tau$, the basket decay rate $r_b$, the additional basket denoted as $\boldsymbol{v_{b(\tau+1)}}$, we update the group vector $\boldsymbol{v_{gk}}$ by applying the incremental rule derived in \autoref{eq: inc_dsa}. 
\begin{align}
    \boldsymbol{v_{gk}^\prime} &= \frac{r_b^{\tau}\boldsymbol{v_{b1}} + r_b^{\tau-1}\boldsymbol{v_{b2}} + \cdots + r_b\boldsymbol{v_{b\tau}} + \boldsymbol{v_{b(\tau+1)}}  }{\tau+1}  \nonumber\\
    &= \frac{\tau r_b\boldsymbol{v_{gk}} + \boldsymbol{v_{b(n+1)}}}{\tau+1} \label{eq: inc_group_vector_rule_2}
\end{align}

Analogous to the previous scenario, we only need access to the current group vector $\boldsymbol{v_{gk}}$ and the number of baskets in the last group $\tau$ in order to obtain the updated last group vector efficiently in constant time. Finally, with the updated group vector $\boldsymbol{v_{gk}^\prime}$, the target user vector after addition $\boldsymbol{v_{u}^\prime}$ is recomputed with access to the last user vector $\boldsymbol{v_{u}}$,  the last group vector $\boldsymbol{v_{gk}}$ and the number of groups $k$ by applying the in-place update rule in \autoref{eq: inplace_dsa}.
\begin{align} 
    \boldsymbol{v_{u}^\prime} &= \frac{ r_g^{k-1}\boldsymbol{v_{g1}} + r_g^{k-2}\boldsymbol{v_{g2}} + \cdots + r_g\boldsymbol{v_{g(k-1)}} + \boldsymbol{v_{gk}^\prime}}{k} \nonumber\\
    &=  \boldsymbol{v_{u}} + \frac{ \boldsymbol{v_{gk}^\prime} - \boldsymbol{v_{gk}}}{k}
    \label{eq: inc_user_vector_rule_2}
\end{align}

\subsection{Decrementally Removing Existing Baskets and Items}
\label{sec:decrementalapproach}
First, we discuss how a user vector will change if a user requests to delete a specific basket from her basket history. In practical applications, this could happen when a customer on an e-commerce website requests to delete a shopping session for privacy reasons or when a customer made a return. It is sufficient just to look at one basket because we can handle the removal of multiple baskets by applying our algorithm repeatedly. Subsequently, we discuss how to remove a single item from a basket instead of the basket as a whole, which corresponds to a scenario where a customer requests to remove certain sensitive items from a previous shopping cart. Again, we can handle these cases via the update rules discussed in \autoref{sec:das}.

\header{Varying group size}
Before diving into the algorithmic details, we first discuss a relaxation of the original algorithm to which we refer as \textit{varying group size}. Given a user's historical baskets, \textit{if we were to maintain a fixed group size}, removing a basket at any location would cause cascading changes in terms of group composition for any subsequent groups following the deleted basket. For users with a large number of baskets, it quickly becomes impractical to regroup and recalculate all group vectors, especially when the basket under deletion is at the beginning. As a consequence, we propose an alternative to avoid expensive regrouping while still maintaining the temporal patterns with recency effects. We allow for a \textit{varying group size}, and thereby restrict the regrouping impact to the enclosing group of the to-be-deleted basket. We demonstrate how the group composition changes under a \textit{varying group size} as follows. 
\begin{align*}
        &[\overbrace{\boldsymbol{v_{b1}}, \boldsymbol{v_{b2}}}^{\boldsymbol{v_{g1}}}, \overbrace{\boldsymbol{v_{b3}}, \boldsymbol{v_{b4}}}^{\boldsymbol{v_{g2}}}, \overbrace{\boldsymbol{v_{b5}}, \boldsymbol{v_{b6}}}^{\boldsymbol{v_{g3}}}, \overbrace{\boldsymbol{v_{b7}}, \boldsymbol{v_{b8}}}^{\boldsymbol{v_{g4}}}]
        \xRightarrow[]{ - \boldsymbol{v_{b3}}} \nonumber\\
        &[\overbrace{\boldsymbol{v_{b1}}, \boldsymbol{v_{b2}}}^{\boldsymbol{v_{g1}}}, \overbrace{\boldsymbol{v_{b4}}}^{\boldsymbol{v_{g2}^\prime}}, \overbrace{\boldsymbol{v_{b5}}, \boldsymbol{v_{b6}}}^{\boldsymbol{v_{g3}}}, \overbrace{\boldsymbol{v_{b7}}, \boldsymbol{v_{b8}}}^{\boldsymbol{v_{g4}}}]
        \xRightarrow[]{ - \boldsymbol{v_{b6}}} \nonumber\\
        &[\overbrace{\boldsymbol{v_{b1}}, \boldsymbol{v_{b2}}}^{\boldsymbol{v_{g1}}}, \overbrace{\boldsymbol{v_{b4}}}^{\boldsymbol{v_{g2}^\prime}}, \overbrace{\boldsymbol{v_{b5}}}^{\boldsymbol{v_{g3}^\prime}}, \overbrace{\boldsymbol{v_{b7}}, \boldsymbol{v_{b8}}}^{\boldsymbol{v_{g4}}}]
\end{align*}
where we show that after removing baskets for $\boldsymbol{v_{b3}}$ and $\boldsymbol{v_{b6}}$, the impacted groups are limited at minimal to $\boldsymbol{v_{g2}^\prime}$ and $\boldsymbol{v_{g3}^\prime}$ respectively.
We differentiate three different deletion scenarios:

\header{Scenario 1: Deleting a basket from an existing group}
If the enclosing group contains more than one basket, deleting a basket from it leads to an updated group vector. Given the to-be-deleted basket vector denoted as $\boldsymbol{v_{bi}}$, the current group vector $\boldsymbol{v_{gj}}$, the basket decay rate $r_b$ and the number of baskets in the enclosing group $\tau$, we can apply the decremental rule (\autoref{eq: dec_dsa}) to obtain the updated group vector,
\begin{align}
\boldsymbol{v_{gj}^\prime} = \frac{ \tau \boldsymbol{v_{gj}} + \mathcal{D}([\boldsymbol{v_{bi}}, ,\cdots,\boldsymbol{v_{b\tau}}])^T \mathcal{R}(r_b, \tau-i)}{(\tau-1)r_b}
\label{eq: dec_group_vector_update_rule_1}
\end{align}
where we have $\mathcal{D}([\boldsymbol{v_{bi}}, ,\cdots,\boldsymbol{v_{b\tau}}]) = [\boldsymbol{v_{b(i+1)}} - \boldsymbol{v_{bi}}, \cdots,\boldsymbol{v_{b\tau}} - \boldsymbol{v_{b(\tau-1)}}, -\boldsymbol{v_{b\tau}}]^T$ and $\mathcal{R}(r_b,\tau-i) = [r_b^{\tau-i}, \cdots, r_b, 1]^T$. Similar to the observations for the decremental rule (\autoref{eq: dec_dsa}), we highlight one important property of \autoref{eq: dec_group_vector_update_rule_1}: we only need to access a slice starting from the deleted basket instead of the full history to perform a deletion. Compared to a full retraining, we reduce the amortized number of baskets impacted by a deletion from $|\mathcal{H}|$ to $|\mathcal{H}|/2$ (assuming a uniform distribution of deletes). Finally, following the in-place update rule (\autoref{eq: inplace_dsa}), we compute the updated user vector $\boldsymbol{v_{u}^\prime}$ from the updated group vector.
\begin{align} 
    \boldsymbol{v_{u}^\prime}
    = \boldsymbol{v_{u}} +  \frac{r_g^{k-i}(\boldsymbol{v_{gj}^\prime} - \boldsymbol{v_{gj}})}{k} \label{eq: dec_user_vector_update_rule_1}
\end{align}


\header{Scenario 2: Deleting a single-basket group} We consider the scenario where a group with a single basket is removed, and the corresponding group vector $\boldsymbol{v_{g_i}}$ will vanish, so we only need to update the user vector $\boldsymbol{v_u}$. Given the number of groups $k$, the vanished group vector $\boldsymbol{v_{g_i}}$, the group decay rate $r_g$, the current user vector $\boldsymbol{v_u}$, we update the user vector by applying the decremental rule (\autoref{eq: dec_dsa}) again, where we have $ \mathcal{D}([\boldsymbol{v_{gi}}, ,\cdots,\boldsymbol{v_{gk}}]) = [\boldsymbol{v_{g(i+1)}} - \boldsymbol{v_{gi}}, \cdots,\boldsymbol{v_{gk}} - \boldsymbol{v_{g(k-1)}}, -\boldsymbol{v_{bk}}]^T$ and $\mathcal{R}(r_g,k-i) = [r_g^{k-i}, \cdots, r_g, 1]^T$.
\begin{align} 
    \boldsymbol{v_u^\prime}
    = \frac{ k\boldsymbol{v_u} + \mathcal{D}([\boldsymbol{v_{gi}}, ,\cdots,\boldsymbol{v_{gk}}])^T \mathcal{R}(r_g, k-i)}{(k-1)r_g}
    \label{eq: dec_user_vector_update_rule_2}
\end{align}

\header{Scenario 3: Removing a single item from a basket}
Instead of requesting to delete a whole basket, one can also delete an item within a basket, and update the corresponding user vector. Deleting an item for a target basket vector $\boldsymbol{v_{bi}}$ will either 1) result in an updated basket vector by changing the corresponding item dimension from one to zero in the basket vector or 2) cause the basket vector to vanish if it only contained the item to remove. In the first case, given the updated basket vector $\boldsymbol{v_{bi}^\prime}$, the current group vector $\boldsymbol{v_{g}}$, the number of baskets in the group $\tau$ and decay rate $r_b$, we update the group vector by applying the in-place update rule (\autoref{eq: inplace_dsa}) again.
\begin{align} 
    \boldsymbol{v_{g}^\prime} 
    &= \boldsymbol{v_{g}} +  \frac{r_b^{\tau-i}(\boldsymbol{v_{bi}^\prime} - \boldsymbol{v_{bi}})}{\tau} \label{eq: dec_group_vector_update_rule_2}
\end{align}
Now that we have $ \boldsymbol{v_{g}^\prime}$, we can apply \autoref{eq: dec_user_vector_update_rule_1} to obtain the updated user vector. In the second case where the target basket vanishes, we fall back to the previous scenario 1 or 2 depending on if the vanished basket also causes a vanished group.


\section{Implementation}
\label{sec:system_overview}

In the following, we discuss a data-parallel implementation of our approach in a popular stream processing system. We briefly show the system overview, then discuss how to leverage the parallel nature of TIFU-kNN, and how to combine incremental and decremental updates to fulfill low-latency requirements. Our implementation is conducted in \textit{Scala} in Apache Spark's Structured Streaming abstraction~\cite{zaharia2016apachespark}. Spark Structured Streaming allows us to express a streaming computation as a state update based on data changes. The Spark engine takes care of running it incrementally and continuously, and of updating the result as new streaming data continues to arrive.

\begin{figure}
    \centering
    \includegraphics[width=\columnwidth]{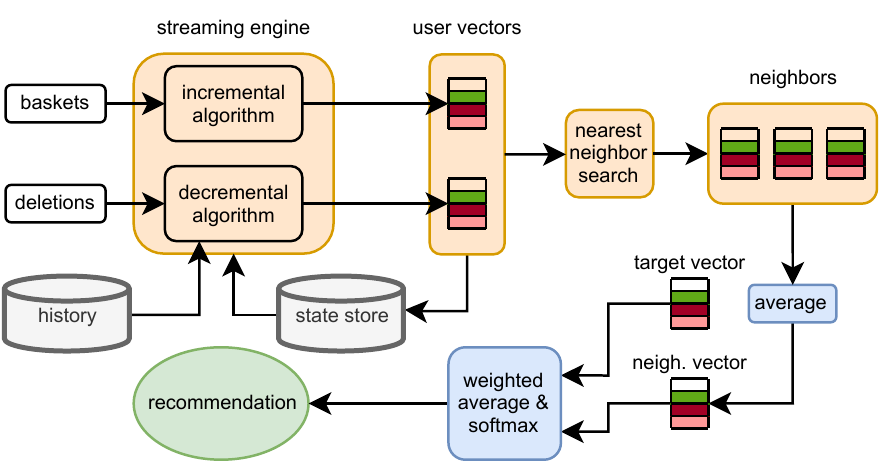}
    \caption{System overview. The streaming engine consists of the incremental algorithm and decremental algorithm which process the newly added baskets and execute deletion requests respectively; the state store maintains the user vectors and provides IO to the streaming engine; next is the nearest neighbor search component which calculates the neighborhood of the target user; finally, the target user vector and neighborhood vector are combined to make recommendations.}
    \Description{System overview for amnesiac recommender systems.}
    \label{fig:systemdiagram}
\end{figure}

\header{System overview} The system (\autoref{fig:systemdiagram}) consists of two stages: 1) the training phase where user baskets and deletion requests are processed and user vectors are produced or updated; 2) the prediction phase where the target user vector and its neighborhood vectors are combined to make recommendations.

\header{Parallelism} Since each user vector is calculated independently, we can leverage this fact to achieve parallelism on the user level, which in turn makes our implementation scale to a large number of users for practical usage. We leverage the \verb#mapGroupsWithState# API \footnote{\url{https://spark.apache.org/docs/latest/structured-streaming-programming-guide.html}} in Apache Spark for this purpose, because this operation allows us to apply user-defined code on grouped datasets to update user-defined state. We treat each user's basket stream as a dataset grouped by a unique user identifier, and the user vectors are stored in a state store.

\header{Joint incremental and decremental state updates} The incremental algorithm and the decremental algorithm in the streaming engine are jointly responsible for updating the user vectors while the former processes incoming user data and the latter handles deletion requests. We show a (simplified) version in \autoref{alg:joint} where the $f_{incr}$ (\autoref{sec:incrementalapproach}) updates the user vector state without accessing the history; the $f_{decr}$ procedure (\autoref{sec:decrementalapproach}) receives the deletion request, the current state and the user's historical baskets and returns the updated user vector state. In our implementation, the users' historical baskets are stored in memory indexed with user identifiers for fast retrieval.  Our code is available under an open source license at \url{https://github.com/0xeeff/amnesiac_recsys}.

\begin{algorithm}
\caption{Joint incremental and decremental state updates}\label{alg:joint}
    \begin{algorithmic}
\Function{updateState}{$currentState$, $input$}
    \State \textit{Input:} the current user vector and the corresponding update (either a deletion request or an additional user basket)
    \State \textit{Output:} the updated state
    \State \textit{Data:} the user historical baskets $\mathcal{H}$ (required by the deletion procedure)
    
    \If {$input.isDeletion$}
        \State $newState \gets$ \Call{$f_{decr}$}{$currentState$, $input$, $\mathcal{H}$}
    \Else
        \State $newState \gets$ \Call{$f_{incr}$}{$currentState$, $input$}
    \EndIf
    \State \Return $newState$
\EndFunction
\end{algorithmic}
\end{algorithm}


\section{Evaluation}
\label{section:evaluation}

We first perform an evaluation using offline ranking metrics, namely, recall and \textit{Normalized Discounted Accumulative Gain} (NDCG) on real-life datasets to show that our approach achieves a comparable performance to our baseline approach of retraining the model from scratch (\autoref{sec:background}). Next, we examine the efficiency and scalability of our approach by evaluating the time to update a user vector or model.

\subsection{Predictive Performance under Incremental/Decremental Updates}
\label{sec:offlinemetrics}

\header{Datasets} The state-of-the-art performance of TIFU-kNN has already been shown by the original authors in \cite{tifuknn}. We only adapt this algorithm and show that our incremental and decremental algorithms do not introduce major performance regressions while at the same time providing for efficient incremental learning and data deletion capabilities. We base our experiments on the publicly available datasets (TaFeng\footnote{\url{https://www.kaggle.com/chiranjivdas09/ta-feng-grocery-dataset}}, Instacart\footnote{\url{https://www.kaggle.com/c/instacart-market-basket-analysis/data}} and ValuedShopper\footnote{\url{https://www.kaggle.com/c/acquire-valued-shoppers-challenge/data}}) and the tuned hyper-parameters ( listed in \autoref{tab:datasetstats}) used for experimentation in the original TIFU-kNN paper~\cite{tifuknn}. The basic statistics for the datasets are shown in \autoref{tab:datasetstats}. These datasets contain timestamped transactions in which items bought in the same order are treated as a basket.



\begin{table*}
    \centering
    \caption[Dataset basic statistics]{Dataset basic statistics and hyper-parameters (from left to right: group size $m$, decay rates $r_b$ and $r_g$, the number of neighbors $k$, the combining weight of the target user vector and neighborhood component $\alpha$).}
    \begin{tabular}{ccccccc}
        Data &  \#users & \#items &\#baskets & avg basket size & avg \#baskets/user &hyper-parameters \\
        \hline
        TaFeng & 13949 & 11997 &79423 & 6.2 & 5.7 & $[7,0.9,0.7,300,0.7]$ \\
        Instacart & 19935 & 7999 &158933 &8.9 & 8.0 & $[3,0.9, 0.7, 900, 0.9]$ \\
        ValuedShopper & 10000 & 7874 &568573&9.1 & 56.9 & $[7, 1.0 , 0.6 , 300 , 0.7]$  \\
        \hline
    \end{tabular}
    \label{tab:datasetstats}
\end{table*}



\header{Experimental setup} For each user, the last basket is held out as a test basket for evaluation. We report the average over all users for the metrics. The baseline, incremental and decremental experiments are implemented according to \autoref{sec:background}, \autoref{sec:incrementalapproach} and \autoref{sec:decrementalapproach} respectively. Recall that the baseline triggers a full retraining for any incremental or decremental updates. For the decremental experiment, as it is difficult to estimate a realistic number of deletion requests, we opt for the setup according to \cite{schelter2021hedgecut}: approximately 1 out of 10,000 users might issue deletion requests in industrial scenarios. Due to the limited number of users in our datasets (\autoref{tab:datasetstats}), we opt for a more extreme estimate of 1 out of 1,000 users, which is a more difficult experimental setting. We select these users at random and delete 10\% of baskets for those users.

\begin{table}
    \centering
    \caption{Predictive performance under incremental and decremental updates compared to the baseline approach of retraining the model from scratch. Our online algorithms provide equal predictive performance.}        
    \begin{tabular}{ccccccc}
        Dataset &  metric & baseline & incr.  & decr.\\
        \hline
        \multirow{4}{*}{TaFeng} 
        & Recall@10 & 0.1298&0.1298& 0.1297  \\
        & NDCG@10 & 0.0847 &0.0847& 0.0847 \\
        & Recall@20 & 0.1859&0.1859  & 0.1856\\
        & NDCG@20 & 0.0816 &0.0816 & 0.0817\\
        \hline
        \multirow{4}{*}{Instacart}
        & Recall@10 & 0.1709 &0.1709 & 0.1710 \\
        & NDCG@10 & 0.5789 &0.5789&0.5788 \\
        & Recall@20 & 0.2291 &0.2291&0.2291  \\
        & NDCG@20 & 0.5037 &0.5037 &0.5035 \\
        \hline
        \multirow{4}{*}{ValuedShopper}
        & Recall@10 & 0.1436&0.1436 & 0.1437 \\
        & NDCG@10 & 0.4058 &0.4058 & 0.4060\\
        & Recall@20 & 0.2133&0.2133& 0.2131\\
        & NDCG@20 & 0.3465&0.3465  & 0.3465\\
        \hline
    \end{tabular}
    \label{tab:offlinemetrics}
\end{table}


\header{Results and discussion} As shown in \autoref{tab:offlinemetrics}, the incremental updates produce exactly the same results and therefore performance metrics as the baseline, which confirms our analysis in \autoref{sec:incrementalapproach}. We expect this behavior, as the incremental updates compute the exact user vectors or models given by the baseline approach. We also observe that the decremental algorithm has no significant negative impact on recommendation performance in terms of NDCG and recall metrics. This observation is confirmed by a t-test on metrics reported for all three datasets. As some might see, decremental updates sometimes even boost the performance metric slightly such as Recall@10 for Instacart and NDDG@10 for ValuedShopper. This is possible because decremental updates could remove noise in historical data hence improve predictive performance for certain users.

\subsection{Update Efficiency}
\label{sec:runtime}

Next, we focus on the runtime performance and scalability of our approach. We demonstrate that we are able to update the model with low latency in constant time per update in the incremental case, and in linear time per update in the decremental case per update.

\header{Experimental setup} Without loss of generality, we assume that we are working with a single user and a vocabulary of one single item. Thus, the basket history looks like $[\{1\},\{1\}, \cdots]$. This simplified view enables us to gain clearer insights into the efficiency of our approach. This setting applies throughout this section.

\begin{figure*}[htbp!]
     \centering
     ~
    \begin{subfigure}[t]{0.31\textwidth}
         \centering
        \includegraphics[width=\textwidth]{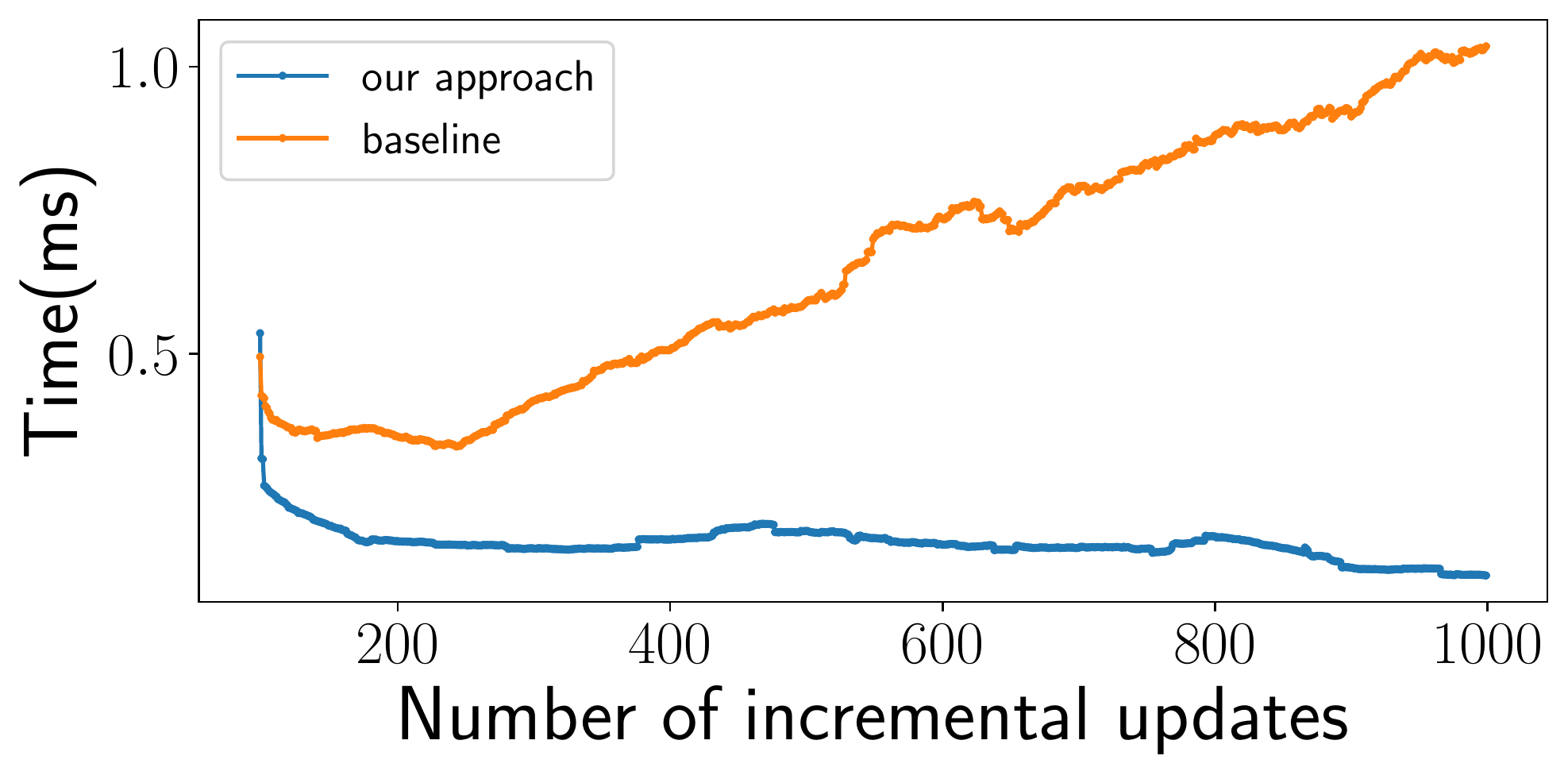}
    \caption{Update times under accumulated basket additions.}
    \label{fig:incremental_compare}
     \end{subfigure}
     \hfill
     \begin{subfigure}[t]{0.31\textwidth}
         \centering
             \includegraphics[width=\textwidth]{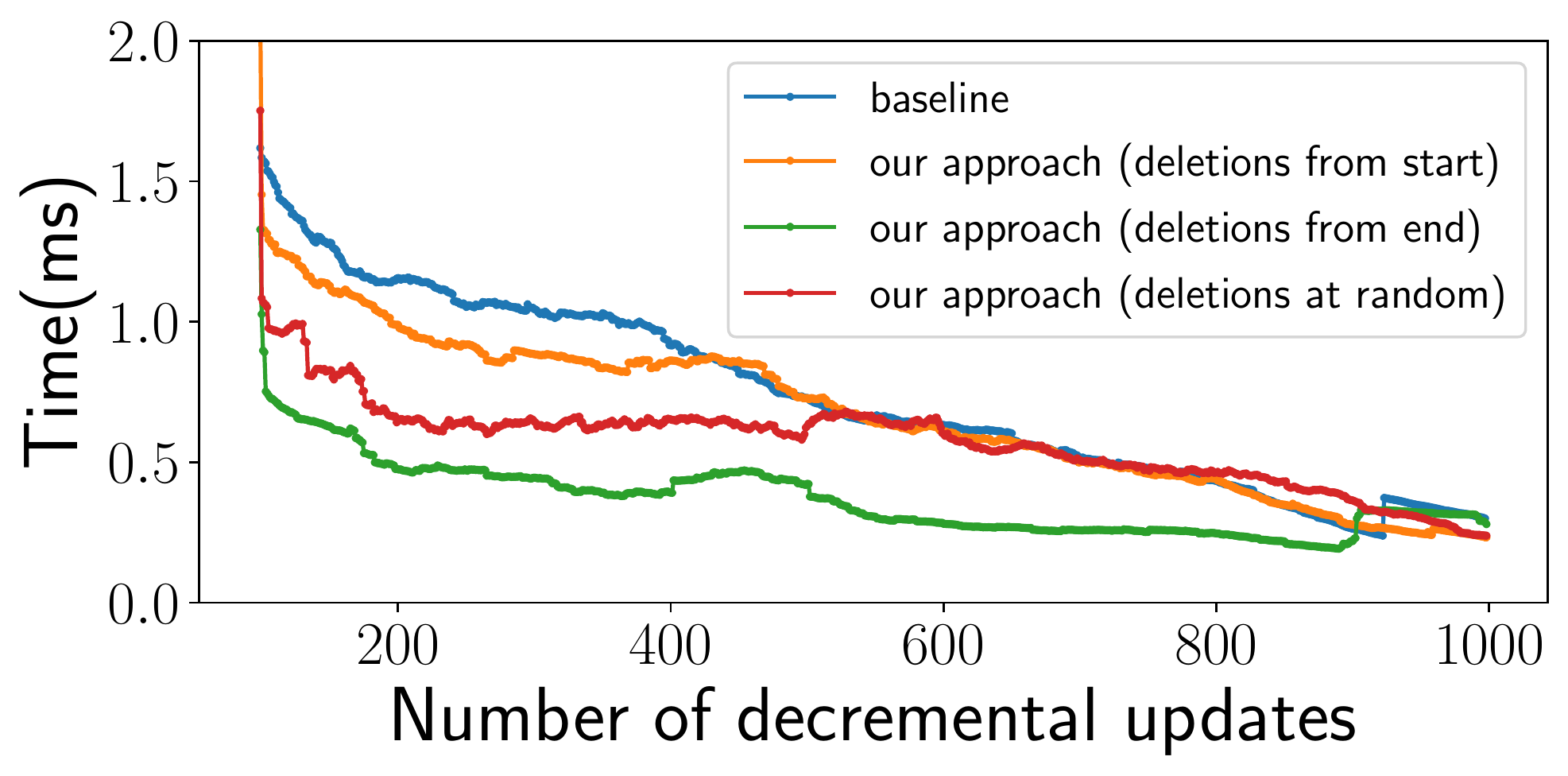}
    \caption{Update times under accumulated basket deletions.}
    \label{fig:decremental_compare_end}
     \end{subfigure}
     \hfill
     \begin{subfigure}[t]{0.31\textwidth}
         \centering
             \includegraphics[width=\textwidth]{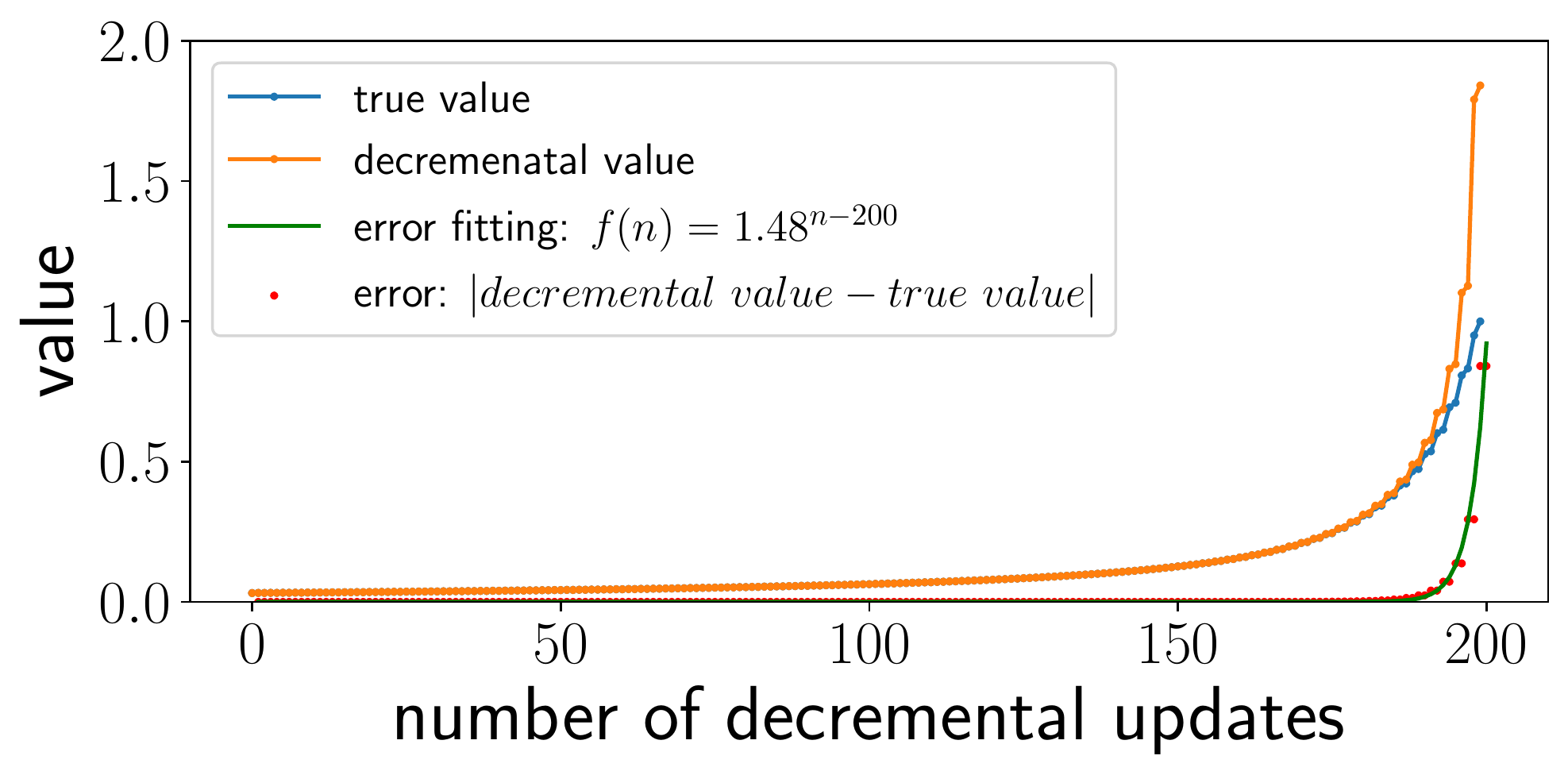}
    \caption{Numeric error caused by a large number of repeated decremental updates (deletions).}
    \label{fig:dec_error}
     \end{subfigure}
        \caption{Updates times of the baseline algorithm (retraining from scratch) and our proposed approach, under accumulated incremental (\autoref{fig:incremental_compare}) and decremental (\autoref{fig:decremental_compare_end}) updates. Our incremental update approach achieves constant time updates regardless of the number of accumulated updates. Our decremental update approach performs consistently better than the baseline: in the best case (deletions from end), it achieves near-constant time efficiency; in the worst case (deletions from start), it performs on par with the baseline; in the canonical case (deletions at random), it shows a performance between the best and worst cases. The overall latency for updates is as low as $1ms$ for the canonical decremental case.}
        \label{fig:baseline_eval}
\end{figure*}


\header{Basket additions} The baseline algorithm needs to access the full history for each update, whereas in our incremental algorithm we only need to access the state and the newly added item or basket once, which can be done in constant time, independent of the history size. In this experiment, we follow the setup where we sequentially add baskets and record the update time along the way. As shown in \autoref{fig:incremental_compare}, our incremental algorithm achieves constant update time efficiency, i.e., the time it takes to retrain does not depend on the size of history, whereas the baseline approach suffers from a linear increase in update time as the number of baskets grows.


\header{Basket deletions} The update time for a deletion depends on the location of the deleted basket. The more recent (closer to the end) it is, the less time it takes to retrain the model. To better illustrate this, we investigate three situations: 1) removing baskets one by one starting from the end; 2) removing baskets one by one starting from the very beginning; 3) removing baskets one by one from any random location. In situation (1), the expected update time should be close to constant. As we remove baskets from the end of the historical baskets, the number of subsequent baskets it needs to access to update the user vector is constant. As we can see in \autoref{fig:decremental_compare_end}, the experiment result validates our assumption though it shows a slight downward trend, which we attribute to regularly triggered maintenance operations in Spark. In situation (2), the expected update time is approximately linear, as we need to access the whole history every time a deletion occurs from the begging of the history. In situation (3), which is a more realistic scenario, the update time lies in between the previous two situations. In summary, our approach is on par with the baseline for worst-case scenarios and consistently outperforms it in other cases.

\subsection{Limitations of Repeated Deletions}
\label{section:limitations}

In the following, we discuss a limitation of our approach, based on a potential numerical instability issue that could occur under extreme conditions (a large number of repeated deletions for a specific user). We theoretically analyze the numerical properties of our decremental algorithm, and empirically a potential issue in an extreme case scenario. 

We can rewrite the user vector update rule (\autoref{eq: dec_user_vector_update_rule_2}) as $ \boldsymbol{u}^\prime = \alpha \boldsymbol{u} + C$ where $\alpha = \frac{k}{(k-1)r_g} > \frac{1}{r_g}>1$. Suppose the initial user vector is $u_o = \hat{u}_0 + \epsilon$, where we use the hat notion $\hat{u}_i$ to represent the true value,  and the $\epsilon$ denotes the initial error.
After $n$ steps of deletions, the accumulated error becomes $|u_n -  \hat{u}_n|= \epsilon   \alpha^n$. Since $\alpha > 1$, the decremental algorithm is numerically unstable. This conclusion implies that there is a limit in terms of how many deletions we can perform for a given user before the error becomes unbearable.

\header{Experimental setup} In this experiment, we first build up the user vector through incremental updates, then we continuously conduct decremental updates during which we calculate the true value of the user vector using the baseline algorithm. The hyper-parameters are fixed: $m=2$, $r_g=0.7$, $r_b=0.9$.


\header{Results and discussion} As shown in \autoref{fig:dec_error}, the error grows exponentially with the number of decremental updates, which is confirmed by the error fitting curve. While this issue might be theoretically worrying, we argue that its impact in practice may be negligible, as the overall number of deletions is typically low, and these deletions will be interleaved with incremental updates. The incremental algorithm is a numerically stable algorithm that helps deduce the accumulated error. In our experiment, to incur 1\% relative error, it takes 180 continuous deletions, which is arguably an extreme case.


\section{Related Work}
\label{section:relatedwork}

\header{Decremental learning} Contrary to conventional machine learning tasks, decremental learning attempts to remove the impact of training data from the trained model efficiently. This line of research dates back to some early work on incremental and decremental versions of Support Vector Machines \cite{svmdecauwenberghs2001incremental, karasuyama2009multipleincrementaldecSVM}, while it has recently gained traction due to the rising concerns for privacy in ML models. Ginart et al. \cite{ginart2019makingaiforgetyou} study data deletion mechanisms for variations of k-means clustering. Schelter et al. \cite{Schelter2020AmnesiaM} propose decremental learning methods for a selection of non-iterative ML models such as item-based collaborative filtering and KNN. Recent work from Guo et al. \cite{guo2020certified} studies this problem for neural networks and shows promising theoretical results albeit apparent scalability bottlenecks due to the requirement of reversing Hessian matrix. Bourtoule et al. \cite{machineunlearning2019} propose a framework in which training samples are divided into shards and trained in silo to obtain a combination of weaker learners, and later on, combined to form a stronger model. When there is a deletion request concerning specific user data, it's sufficient to update the corresponding impacted weaker learner.



\header{Deep learning for NBR} Deep learning \cite{alexnet,gan,dqn,gpt3,dalle} gains momentum in RS given its potential in learning powerful latent representations of users/items \cite{dlrecsys}. Its ability to model sequential patterns is especially interesting for NBR. Many new methods based on recurrent neural networks and attention mechanisms \cite{transformer} have been proposed \cite{hidasi2016sessionbased,hidasi2018recurrentGRU4RECplus,yu2016dynamic,li2017neural,ying2018sequential,liu2018stamp,ren2019repeatnet}. However, questions for the effectiveness of deep learning in the RS domain, especially next basket recommendation/sequential recommendation, remain open. Other studies show that simple neighborhood-based methods outperform complex neural models easily \cite{jannach2017recurrent, kersbergenlearnings, tifuknn}.




\section{Conclusion}

In this work, we focus on efficient data deletion mechanisms for recommender systems in light of privacy protection and introduce the problem of decremental learning to Next Basket Recommendation. We then present efficient and effective algorithms capable of learning and forgetting with millisecond-level latency. Our approach is evaluated on real-world datasets and demonstrates its effectiveness in terms of ranking metrics (recall and NDCG). Moreover, it also proves to have desirable update time efficiency as shown in our experiments. In addition to that, we provide a fully parallel implementation in Apache Spark to facilitate practical use cases. For theoretical completeness, we also study the limitations under extreme conditions, that is, the decremental algorithm could suffer from numerical instability for a large number of continuous deletions. For future work, we intend to investigate: 1) how to incorporate incremental and decremental algorithms for other more complex recommendation scenarios; 2) numerical stability properties for other ML models under similar settings. In the end, we hope for more advances in this area to realize a future of truly ``amnesiac'' recommender systems.

\header{Acknowledgments} {\em This work was supported by Ahold Delhaize. All content represents the opinion of the authors, which is not necessarily shared or endorsed by their respective employers and/or sponsors.}

\newpage
\bibliographystyle{ACM-Reference-Format}
\bibliography{nbr}
\end{document}